\documentclass{appolb}
\usepackage{graphicx}

\usepackage{bm}
\usepackage{amsfonts}
\usepackage{amsmath}
\usepackage{amssymb}
\usepackage{graphicx}
\newcommand{\Oo}{{\Delta_2}}

\newcommand{\eq}{\begin{eqnarray}}
\newcommand{\eqx}{\end{eqnarray}}
\newcommand{\ba}{\begin{equation}}
\newcommand{\ea}{\end{equation}}
\newcommand{\bal}{\begin{align}} 
\newcommand{\eal}{\end{align}}
\newcommand{\f}[2]{\frac{#1}{#2}}

\newcommand{\n}{\nonumber \\}

\newcommand{\bit}{\begin{itemize}} 
\newcommand{\eit}{\end{itemize}}

\def\la{\label}
\def\nn{\nonumber \\}
\def\bi{\bibitem}

\def\d{\partial}
\def\th{\theta}

\def\lam{\lambda}
\def\al{\alpha}

\def\va{\varphi}

\def\eps{\epsilon}

\def\la{\label}
\begin{document}
 \eqsec  
\title{Positivity and unitarity constraints on dipole gluon distributions.%
\thanks{Presented at ``Particle Production in Hadronic Collisions'',
Meeting in Krakow in honor of professor Andrzej Bialas for his 80$^{th}$
anniversary.}%
}
\author{Robi Peschanski
\address{Institut de Physique Th\'eorique,\\
CEA, IPhT, F-91191 Gif-sur-Yvette, France\\
CNRS, URA 2306}
\\
}
\maketitle
\begin{abstract}
In the high-energy domain, gluon transverse-momentum dependent distributions in nuclei obey constraints coming from positivity and unitarity of the colorless QCD dipole distributions through Fourier-Bessel transformations. Using mathematical properties of Fourier-positive functions, we investigate the nature of these constraints which apply to dipole model building and formulation.
\end{abstract}
  
\section{Introduction}
The QCD dipole formalism \cite{mueller} has proven to be quite successful
as a tool for describing ``low-x physics''. This is the domain of high-energy scattering of particles on nuclei at high energy and moderate but high enough tranverse momentum exchange allowing the QCD coupling constant to be small. One relevant example is the 
description and model building for the {\it transverse momentum dependent} 
(TMD)  gluon distributions in  nuclei. They appear in the formulation of physical
observables such as forward jet production \cite{marquet1} and forward dijet correlations \cite{gluon} off nuclei by scattering of protons. In the QCD dipole formalism, i.e. in 
the large $N_c$ and leading-log approximation of perturbative QCD, they 
are related to the {\it size-dependent} distribution of colorless 
gluon-gluon ($gg$) dipoles in the target nucleus. 
 
Within some simplifying assumptions \cite{marquet1}, this relation takes the form of a pair of Fourier-Bessel transforms, namely
\eq 
{\cal G}(Y,k) &=& \int_0^\infty rdr\  
J_0(kr)\ 
 {\cal S}(Y,r)\ ,
\nn 
{\cal S}(Y,r)\ &=&\  \int_0^\infty  k{dk}\  
J_0(kr)\ 
{\cal G}(Y,k)\ ,
\la{fpair}
\eqx
where $1-{\cal S}(Y,r)={{\cal T}}(Y,r)$ is  the $gg$
 distribution in the target as a function of their size $r$ at total 
c.o.m. energy $e^Y.$ Its Fourier-Bessel partner ${\cal G}(Y,k)$ enters into the expression of  the TMD gluon distribution 
\ba 
xG(x,k)=\frac{k^2N_c}{2\pi^2\al_S}\ A_\bot\ {\cal G}(Y,k)\ , 
\la{TMD}
\ea 
where $x=e^{-Y},$ $\al_S$ is the QCD coupling constant, and $A_\bot$ the 
target transverse area\footnote{This, so called {\it dipole gluon}, 
which distribution is to be distinguished from the Weiszs\"acker-Williams 
gluon distribution. The  Weiszs\"acker-Williams distribution cannot be expressed as the Fourier transform of a QCD dipole distribution \cite{gluon}.}. 

Interestingly enough, the dipole formalism  is submitted to positivity and unitarity conditions which gives rise nontrivial constraints on the pair of Fourier transforms \eqref{fpair}. Let us specify them as follows.

i) {\it Positivity constraint on} $\ {\cal T}(Y,r)$. The dipole gluon distribution $xG(x,k)$ is expected to be positive. In fact it is required to be so, since it is proportional \cite{marquet1,gluon} to physical observables. Hence from \eqref{TMD}, ${\cal G}(Y,k)$ is positive, and through \eqref{fpair} it induces nontrivial mathematical constraints on the $gg$ dipole distribution in the target ${\cal T}(Y,r).$ Its Fourier-Bessel transform should be positive.

ii) {\it Unitarity constraint on} $\ {\cal G}(Y,k).$ The dipole distribution  ${\cal T}(Y,r)$ is also expressed as  
the dipole-target total cross-section at total incident energy $E=e^Y$, up to a normalization. As such 
it obeys the  S-matrix unitarity condition ${\cal S} = 1-{\cal T}(Y,r)\ge 0.$ Hence, through 
relations \eqref{fpair}, unitarity implies that ${\cal G}(Y,k)$ has to 
have a positive  Fourier-Bessel transform.

In both both cases, one has functions whose Fourier transforms are positive (here, 2-dimensional radial i.e. Fourier-Bessel ones). The aim of the present contribution is to show some interesting physical consequences of this mathematical property.  We call it $\cal F$-positivity.

\section{Fourier-positivity}
\la{mathsec}
$\cal F$-positivity is the mathematical property of real-valued functions whose Fourier transforms are positive \cite{maths}.
Contrary to expectation, there is no explicit parametrization of the set of $\cal F$-positive functions. They  instead can be characterized by an infinite set of necessary conditions, which constitute the B\"ochner theorem \cite{bochner}. For a given $\cal F$-positive real function $\psi(\vec v)$ whose d-dimensional Fourier transform $\va(\vec w)$ is positive, the B\"ochner theorem states that the function $\psi$ is {\it positive definite}, that is
\ba
\sum_{i,j=1}^{n}u_i\  {{\psi }(\vec{v_i}\!-\!\vec{v_j})}\  u_j\ >\ 0\ ,\quad 
\forall  u_i,\ \forall  \vec v_i,\
\forall n\ . 
\la{positive}
\ea
In the case of the Fourier-Bessel transforms \eqref{fpair}, the  conditions  \eqref{positive} apply to any set of two-dimensional vectors $\vec v_i.$ Hence, for any $n \in {\mathbb N}$ and for any set of numbers 
$\{{ u_i}, i=1,...,n\},$ the $n \times n$ matrix $\mathbb M$ with 
elements 
${{\psi}(|\vec{v_i}\!-\!\vec{v_j}|)}$ is positive definite. This is equivalent to the property that the lowest eigenvalue of $\mathbb M$
 remains positive for all $\vec v_i, u_i$ and all values of $n.$ In the case of \eqref{fpair}  two-dimensional transverse position coordinates $\vec r$ (with $r= |\vec r|$) or transverse momentum space $\vec k$ (with $k= |\vec k|$) may be involved, 
 depending on the required physical constraint we shall consider later on.

Applying the whole set of consitions \eqref{positive} is not realistic for practical purposes. For this sake, we have developed in the recent years \cite{gipe,newfourier,diraccomb} specific 
tools for practical tests of $\cal F$-positivity. They are formulated, in various forms, in terms of an optimized finite subset of necessary conditions issued from 1- and 2-dimensional versions of the relations \eqref{positive}. 

One of the conditions coming from the B\"ochner theorem which appears to be relevant for our  problems is the following.
Let us consider the $3\times3$ matrix ${\mathbb M}_3$ with matrix elements
 \ba
\{{\mathbb M}_3\}_{i,j}\ \equiv\  
\left\{\psi(|\vec{v_i}\!-\!\vec{v_j}|)
\,\right\}
,\quad \vec {v_i} \ =\ \{0,0\}\ ,\ \{0,v\}\ ,\ \{v\sin\th,v\cos\th\}\ ,
\la{3-lattice}
\ea
 which leads to the $\cal F$-positivity conditions for the matrix
\ba
{\mathbb M}_3=
\begin{pmatrix}
\psi(0)& \psi(v)& \psi(2v\sin \frac\th 2)\\
\psi(v)& \psi(0)& \psi(v)\\
 \psi(2v\sin \frac \th 2)& \psi(v)&\psi(0)\\ 
\end{pmatrix}\ .
\la{3x3matrix}
\ea
Positive-definiteness implies positivity of the matrix determinant
and of its minors along its diagonal. This leads, up to a rescaling of $v,$ to the inequalities
\ba
\psi(0)\ > \  
\psi(v)\  > 
\  2\ \frac {\psi^2\left( v/ [2\sin \frac \th 2]\right)}{\psi(0)}
-\psi(0),\ \forall v > 0, \forall \th \in [0,\pi]\ .
\la{fptest}
\ea
Note that a larger number of points, provided they include the points of \eqref{3-lattice} 
still leads to
condition \eqref{fptest}, together with others, forming a hierarchy of necessary condition  \cite{diraccomb} for $\cal F$-positivity.

An important addendum to the $\cal F$-positivity tests for a 2-dimensional radial function $\psi(v),$ has been noticed in \cite{newfourier}. They can be extended to the action of the radial Laplacian on $\psi(v),$ namely
\eq
\Oo \left[\psi\right](v)\ &\equiv&\ - \f 1v\ \f d{dv}\left(v\ \f {d\psi(v)}{dv} \right)
\n
 &=&  \int_0^\infty w^3dw\  
J_0(vw)\ \va(w)\, > 0\ ,
\la{operatorJ}
\eqx
where $\phi(w)>0$ is  the positive Fourier-Bessel transformed of $\psi(v).$ Hence ${\cal F}$-positivity tests then apply not only to $\psi,$ but also to  $\Oo \left[\psi\right],$ and its iterations, provided the integrals, such as in \eqref{operatorJ} for the first one, remain convergent.

\section{Positivity constraints on the dipole distribution}
\la{positif}
Through the second equation \eqref{fpair}, the positivity of the gluon distribution ${\cal G}(Y,k)$  induces $\cal F$-positivity constraints on the dipole amplitude ${\cal T}(Y,r)=1-{\cal S}(Y,r).$ 

In order to conveniently formulate these $\cal F$-positivity constraints, let us turn to the first equation \eqref{fpair}. By double integration by part on the right hand side, one obtains
\eq 
{\cal G}(Y,k) &=& \int_0^\infty rdr\  
J_0(kr)\ 
 (1-{\cal T}(Y,r))\n
\ &=&\  \int r{dr}\  
J_1(kr)\ 
\frac {\d}{\d r}{\cal T}(Y,r)\ \n
\ &=&\  \int r{dr}\  
J_0(kr)\ 
\f 1r\frac {\d}{\d r}\ \left(r\frac {\d}{\d r}{\cal T}(Y,r)\right)\ \n
\ &=&\  \int r{dr}\  
J_0(kr)\ 
\Oo \left[{\cal S}\right](Y,r)\  >\ 0\ .
\la{fpair0}
\eqx
Performing the integrations, we used the known derivative relations between Bessel functions, successively,   $rJ_0(r) = \frac {\d}{\d r}J_1(r)$ and $J_1(r)=-\frac {\d}{\d r}J_0(r).$ 

Now the key point, as discussed in our recent work \cite{fourierdipoles}, is the behavior of ${\cal T}(Y,r)$ when the dipole size $r\to 0.$ The standard leading order QCD behavior near the origin is given by the property of ``color transparency'',
${\cal T}(Y,r)\propto r^2.$ Following \cite {fourierdipoles}, higher order QCD corrections of the dipole amplitude
near the origin leads to the modified behavior 
\ba
{\cal T}(Y,r)\ \propto\ r^{2+\eps} \quad {\rm when}  \quad r\to 0\ ,
\la{transparency}
\ea
where $0<|\eps| \ll 1$  parameterizes the slight deviations from color transparency. On the one hand,  they are expected to come from the running of the coupling constant. On the other hand, they are generated  by the resummation of the perturbative QCD expansion in the low-x (high $Y$) domain of dipole models. 

Interestingly enough, the ${\cal F}$-positivity inequalities \eqref{fptest} impose the simple but nontrivial condition $\eps \leq 0.$ in \eqref{transparency}. This condition has consequences \cite{fourierdipoles} on various dipole models when the running QCD coupling constant $\al_S(r)$ at short dipole separation is taken into account.

Let us consider  for instance the saturation model consistent with the leading order DGLAP evolution \cite{msat}. The dipole amplitude reads 
\ba
{\cal T}(Y,r) = 1-{\cal S}(Y,r) = 1-
\exp{
\left(
-\frac {\pi^2r^{2}}{3\sigma_0}\ 
\alpha(\mu^2) xg(x,\mu^2)\right)
} ,\ x\equiv e^{-Y}\ .
\la{satdglap}
\ea
Here  $xg(x,\mu^2)$ is the gluon distribution function in the proton considered at 
momentum fraction $x$ with  $r$-dependent momentum scale $\mu^2 =  C/{r^2}+\mu_0^2,$
 with $\alpha(\mu^2)\propto \log \Lambda_{QCD}^2/\mu^2$ 
and $\sigma_0, C$ are phenomenological 
constants fitted to deep-inelastic data. The interest of this model is that it combines the saturation effect ${\cal S}\to 1$ at large dipole size $r$ with a behavior at small $r$ compatible with the DGLAP evolution equation.

In Eq.\eqref{satdglap},  the 
color transparency behavior ${\cal T}(Y,r) \propto r^2,\ r\to 0,$ is naturally obtained for fixed $\al_S$ and constant $xg(x,\mu^2)$
at first order of the QCD perturbative expansion. This is equivalent to  the original Golec-Biernat and W\"usthoff model \cite{golec} which reads
\begin{align}
 {\cal T}(Y,r) =
1-\exp{
\left(
{-\frac {r^2}4 Q^2_S(Y)}
\right)
}\ ,
\la{GBW}
\end{align}
with $Q_S(Y)$ is the ``saturation scale''.  In fact the model  verifies the $\cal F$-positivity constraints, as it is obvious  from \eqref{fpair} by Gaussian integration.

Adding higher orders in the coupling constant modifies that behavior and its consequences. The running of $\al_S(\mu^2)\sim 1/\log(1/r^2)$ leads to an effective value $\eps_{run}<0$ in Eq.\eqref{transparency}. On the other hand, the summation of the double leading logarithms of the QCD perturbative expansion at small $x$ leads to
\ba
xg(x,\mu^2) \approx 
\sum_n 
\frac{\left[\ Y \int_{\mu_0^2}^{\mu^2}\!\! \alpha(k^2)\  {dk^2}/{k^2}\ \right]^n}{(n!)^2} \sim \ 
e^{cst.\sqrt{Y\ \log\log\f{1}{r^2}}}\ .
\la{g}
\ea
 Hence in this case, the overall modification of the color transparency
 behavior due to the increase of the gluon parton distribution function at small $r$ leads to a positive contribution $0<\eps_{resum}\ll 1$ to the behavior \eqref{transparency}. The overall effect of  both higher order contributions leads to 
\begin{align}
\f {{\cal T}(Y,r)}{r^2}\ \approx\ \frac { 
e^{cst.\sqrt{Y\ \log\log{\frac 1{r^2}}}}}{{\log  \frac1{r^2}}}
 \to\ 0\quad {\rm when}\ r\!\to\! 0\ .
\la{limit}
\end{align} 
Hence one finds that the compensation of the running effect by the resummation one in model \eqref{satdglap} is incomplete, leading to since
$\eps = \eps_{run}+\eps_{resum}<0$ in Eq.\eqref{transparency}. The effect may  be  small in absolute value, but is non zero. 

$\cal F$-positivity of the model \eqref{satdglap} is thus violated. Hence the corresponding ${\cal G}(Y,k)$ is not everywhere positive. This can be verified by explicit Fourier transform. The phenomenological and theoretical relevance of such and similar behavior in various dipole models has been discussed in Ref.\cite{fourierdipoles}. It may lead to a reformulation of dipole models in the presence of higher order QCD corrections. We shall discuss this conclusion further on in section \ref{discussion}.

\section{Unitarity constrains on the gluon distribution}

The unitarity condition on the dipole distribution  ${\cal S}(Y,r)=1-{\cal T}(Y,r)\ge 0,$ induces $\cal F$-positivity of the gluon TMD distribution ${\cal G}(Y,k)$
as shown by the relations Eq. \eqref{fpair}.  Indeed,
let us consider  the second line of Eq.\eqref{fpair}. The constraint reads
\bal 
{\cal S}(Y,r)\ =\  \int k{dk}\  
J_0(kr)\ {\cal G}(Y,r)\ \ge\ 0\ ,
\la{fpair1}
\end{align}
which involves the $\cal F$-positivity constraints \eqref{positive} for ${\cal G}(Y,r)$. The necessary condition \eqref{fptest} of section \ref{mathsec} reads
\ba
{\cal G}(Y,k=0)\ > \  
{\cal G}(Y,k)\  > 
\  2\ \frac {{\cal G}^2\left(Y, k/ [2\sin \frac \th 2]\right)}{{\cal G}(Y,0)}
-{\cal G}(Y,k=0)\ .
\la{fptest1}
\ea
The limiting quantity ${\cal G}(Y,k=0)$ is difficult to estimate directly from the gluon transverse momentum spectrum. However, from the first line of \eqref{fpair}, one finds the expression
\eq 
{\cal G}(Y,0) = \int_0^\infty\!\! rdr\  
 {\cal S}(Y,r)\ ,
\la{G0}
\eqx
which can be estimated from the associated dipole model.\ 
In the original GBW model \eqref{GBW}, one finds ${\cal G}(Y,0)= 2\ Q_S^{-2}(Y),$ which  a  characteristic length squared scale of saturation models. More generally, a model verifying  ``geometric scaling'' \cite{geom},   ${\cal S}(Y,r)= {\cal S}(rQ_S(Y))$ gives rise to a similar dependence on the saturation scale, namely
\eq 
{\cal G}(Y,0) = \
 \ Q_S^{-2}(Y)\ \int_0^\infty\!\! \rho d\rho\  
 {\cal S}(\rho)\ = \lambda^{-2}\ Q_S^{-2}(Y)\ ,
\la{G02}
\eqx
where $\lambda = {\cal O} (1)$ is specified by the corresponding dipole model. All in all, for generic saturation models (having only at most small violations of geometric scaling) one  expects  ${\cal G}(Y,0) \sim {\cal O} (Q_S^{-2}(Y)).$

Combining Eqs. \eqref{fptest1} and \eqref{G02}, one obtains
\ba
 1\ >\ {{\tilde g}\left(Y, k\right)}\ > \ 2\ {\tilde g}^{2}\left(Y, \frac k {2\sin \frac \th 2}\right)
-1\ ,
\la{fptest2}
\ea
where we introduced the short-hand notation
\ba
{{\tilde g}\left(Y, k\right)}\ \equiv \frac {{\cal G}\left(Y, k\right)}
{{\cal G}\left(Y, 0\right)}\ =\ \lam^2 Q_S^2(Y)\ {\cal G}\left(Y, k\right)\ .
\la{notation}
\ea

Hence the relation \eqref{fptest2} induces  bounds on the magnitude of the TMD gluon distribution
in the whole momentum range. In particular, from the leftmost inequality in \eqref{fptest2}, it appears that
\ba
{\tilde g}\left(Y, k\right)\equiv \lambda^{2} Q_S^{2}(Y){\cal G}\left(Y, k\right)\ <\ 1\ .
\la{ineqg}
\ea
Thus ${\cal G}$ cannot rise significantly above the inverse squared of the saturation scale. The quantitative $upper$ bound on $\tilde g$ is a function of the constant $\lambda$ and thus of  the dipole model in use.

It is also interesting to take into account the second  inequality of Eq.
\eqref{fptest2}. For a given value $k$ of the transverse momentum and varying the angle $\th \in [0,\pi],$
the corresponding range  is $[k/2,\infty].$ Quoting $\tilde g_{max}(Y,k)\le 1,$ the maximum value of $\tilde g$ in this range, we obtain now a $lower$ bound on $\tilde g.$ Combining the upper and lower bounds, one gets
\ba
1\ >\  \tilde g_{max}(Y,k)\ >\ {{\tilde g}\left(Y, \kappa\right)}\ > \ 2\ {\tilde g_{max}(Y,k)}^{2}-1\quad \kappa\ \in\ [k/2,\infty]\ .
\la{fptest3}
\ea
However, in order for the lower bound in Eq. \eqref{fptest3} to be operating, one has the condition 
\ba
\tilde g_{max}(Y,k)>\sqrt2/2\ ,
\la{condition}
\ea
which limits the range of validity of the lower bound in $k.$

One typical example is when $\tilde g$ is a monotonically decreasing function of $k$ then with $\tilde g_{max}(Y,k)=\tilde g(Y,k/2).$ Then \eqref{condition} translates into 
\ba
1\ >\ {{\tilde g}\left(Y, k\right)}\ > \ 2\ {\tilde g^2(Y,k/2)}-1
\la{fptest4}
\ea
with the condition $\tilde g(Y,k/2)>\sqrt2/2.$ In all cases, Eq.
\eqref{fptest4} works in the lower  $k$ range.

\section{Discussion}
\la{discussion}

There are interesting  phenomenological and theoretical consequences of the mathematical constraints due to positivity and unitarity derived in the previous sections. In particular, the positivity constraints of section  \ref{positif} appear to cast a doubt on some formulations of  QCD dipole models when the running of the QCD coupling constant in coordinate space $\al_S(r)$ is taken into account. Asymptotic freedom for short separation of the gluons in the QCD dipole apparently leads to a contradiction with $\cal F$-positivity.

\begin{figure}[ht]
\centerline{%
\includegraphics[width=8cm]{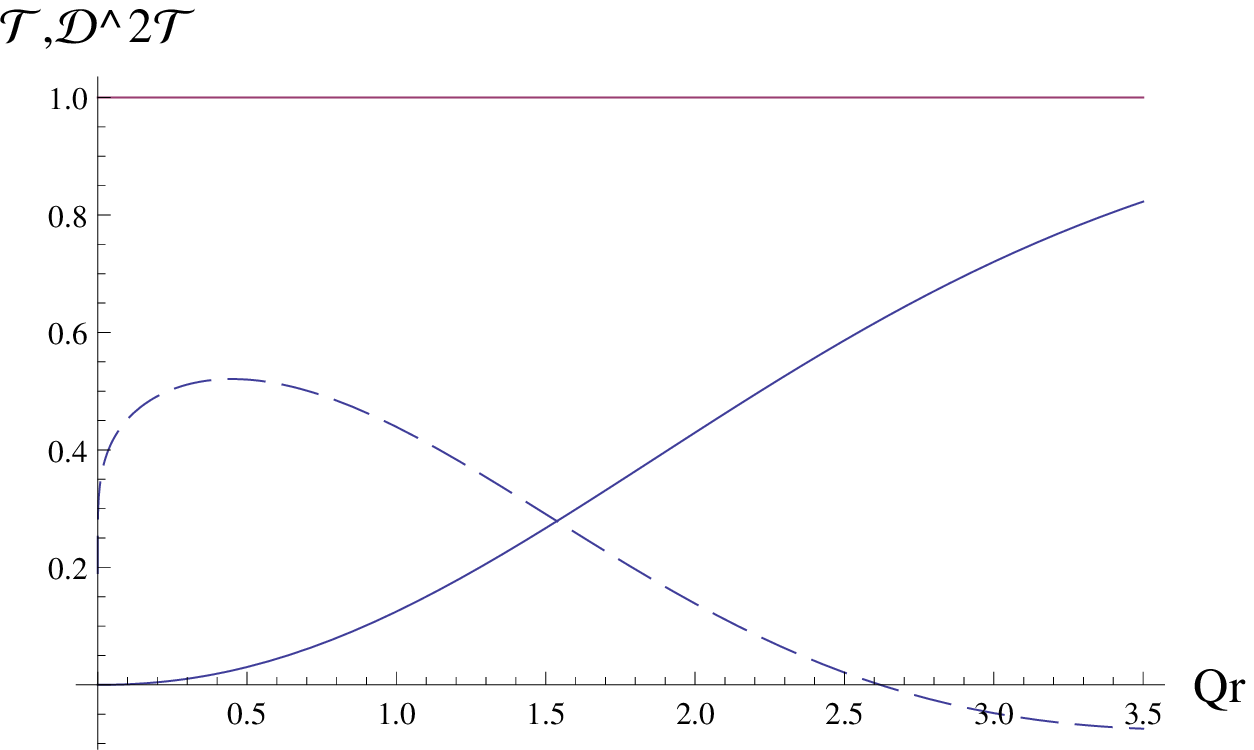}
\includegraphics[width=6cm]{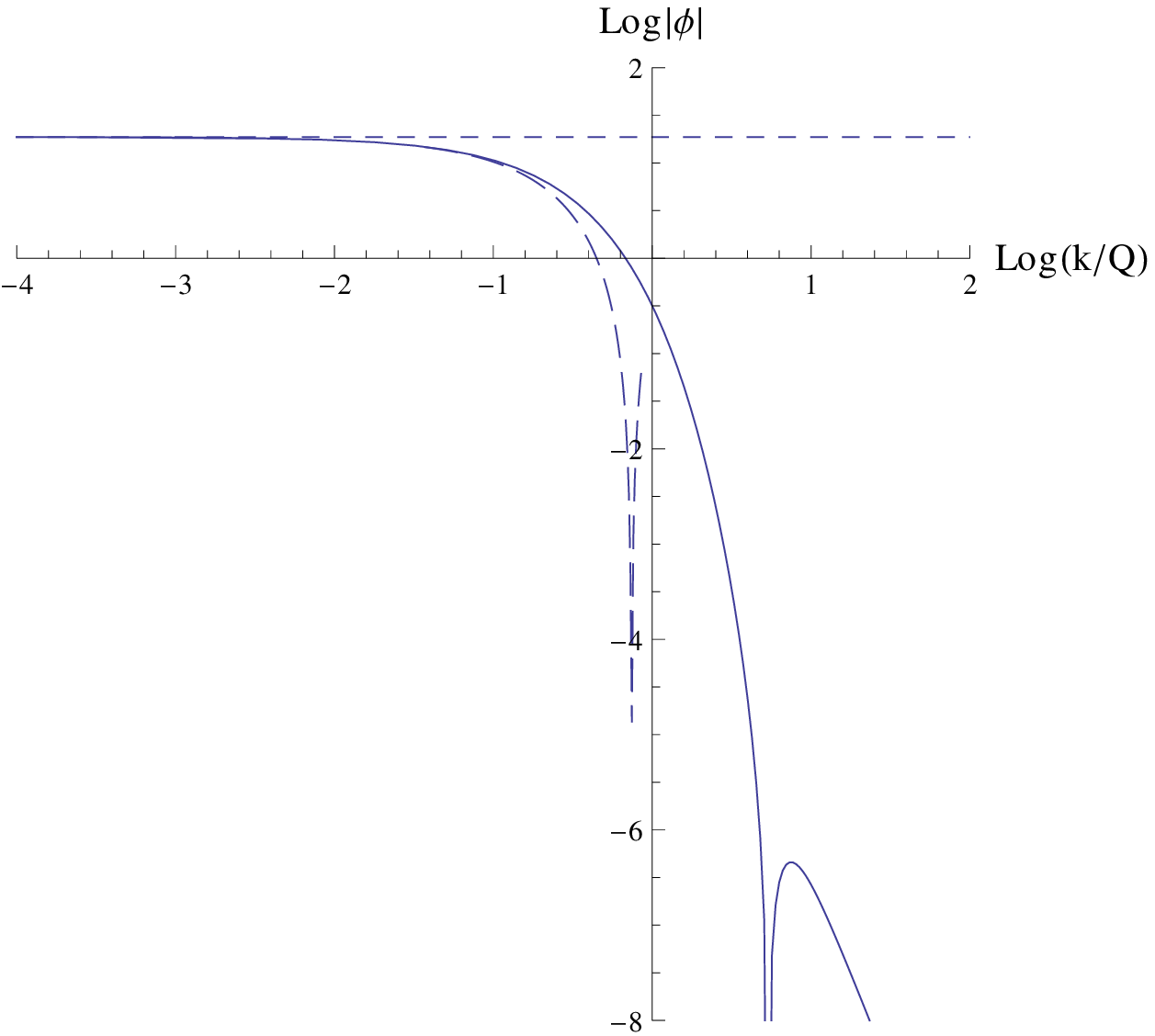}
}
\caption{{\it Positivity and unitarity tests of the McLerran-Venugopalan model with running coupling.}
\newline 
Left, the dipole amplitude:  $full\ line,$ ${\cal T}(Y,r),$ $dashed\ line,$ ${\cal D}^2{\cal T}(Y,r),$ see \eqref{def}, as a function of $rQ.$
 \newline {Right,} its Fourier-transform $\va(Y,k/Q) \equiv 
{\cal G}(Y,k)$ in a $\log |\va|,\log (k/Q)$ plot, and the upper and lower bounds from Eq.\eqref{fptest4} due to unitarity ;
\newline $Full\ line,$ the absolute value of $\va(Y,k/Q) \equiv 
{\cal G}(Y,k),$ showing the positivity violation at a value of $k/Q,$ after the dip signalling the zero. 
The upper and lower bounds due to unitarity are shown with discontinulous lines: $short-dashed\ line,$ the upper bound $\va(Y,0),$ $long-dashed\ line,$ the absolute value of the lower bound. Note that the lower bound becomes negative (and thus not operating) beyond the dip signaling the zero.}
\label{1}
\end{figure}
In order to show the relevance of positivity and unitarity constraints on an example, we show in figure \ref{1} the results for the McLerran-Venugopalan (MV) model with running coupling \cite{MV}, using phenomenologically realistic rapidity and model parameters. For concreteness, we thus choose  a specific  version used in a recent summation model \cite{iancu}. 

The corresponding dipole amplitude reads
\begin{align}
 {\cal T}(Y,r) = \left\{
1-\exp{
\left[
-\left(
{\frac 14}(rQ(Y))^{2}
{\alpha_S(rC)}
\left[
1+\log\left(\frac{\bar \alpha_{sat}}{{\bar \alpha_S(rC)}}\right)
\right]
\right)^{ p}
\right]}
\right\}^{{1/p}}\ ,
\la{rcMV}
\end{align}
where ${\alpha_S(rC)}\propto {\log^{-1}(1/(rC)}$ is the running QCD 
coupling in transverse coordinate space and $C,\alpha_{sat},p$ are 
phenomenological constants. 
Following Eq.\eqref{fpair0},  ${\cal F}$-positivity constrains  the double derivative of the dipole amplitude \eqref{rcMV}
defined as follows
\ba
{\cal D}^2{\cal T}(Y,r)\equiv\f 1r\frac {\d}{\d r}\ \left(r\frac {\d}{\d r}{\cal T}(Y,r)\right)
\la{def}\ .
\ea

i) {\it Positivity constraints.} The positivity violation and its source appear clearly on the curves drawn in Fig.\ref{1}. The figure on the left side shows both the amplitude $\cal T$ and its double derivative ${\cal D}^2{\cal T},$ Eq.  \eqref{def}. This last function has a positive derivative near the origin (i.e. $\eps > 0$ in Eq.\eqref{transparency}) which violates the ${\cal F}$-positivity constraint \eqref{fptest}. On the right hand side, the positivity violation is made manifest 
by explicit computation of the Fourier-Bessel transform. 
The positivity violation can be traced back to the inverse logarithmic coupling of the running constant which is not compensated by the log-log term which appears into brackets in \eqref{rcMV}.
Hence, the formulation of QCD dipole models  with a $r$-dependent coupling constant, $\al_S(r),$  leads to a violation of the expected positivity of the TMD gluon distribution. This violation appears on the $ultra\!-\!violet$ side (larger $k/Q$) of the TMD gluon distribution spectrum.

ii) {\it Unitarity constraints.} Fig.\ref{1}, left, shows explicitely that the unitarity constraints \eqref{fptest4} are satisfied by the amplitude of the McLerran-Venugopalan model with running coupling, as can be easily checked on Eq.\eqref{rcMV}. The constraints are shown by discontinuous lines in Fig.\ref{1}, right. They are clearly operating  on the $infra\!-\!red$ (lower  $k/Q$) behavior of the gluon TMD distributions. This comes from the upper bound being the limiting value ${\cal G}$ at $k=0,$ and from the limiting condition \eqref{condition} on the lower bound to be valid.

As a concluding remark, both positivity and unitarity constraints provide quite general constraints on the QCD dipole models, which are of common and useful use in high energy phenomenology. We see that even the application of the B\"ochner theorem on the low rank $3\times 3$ matrix case, leads to nontrivial consequences. These are remarkably distributed between the ultra-violet (for positivity) and infra-red (for unitarity) sectors of the TMD gluon spectrum.

Our study was limited to leading-log orders of dipole models and to the lower rank matrix tests of the B\"ochner theorem. As an outlook, it is clear that our observations ask for  a developed study of ${\cal F}$-positivity constraints beyond leading orders and smaller matrix rank.

\section{Acknowledgements}
The studies presented here mainly come from a thorough and long term collaboration with Bertrand Giraud (IPhT, Saclay) and  I warmly thank him for this fruitful common work.

The present contribution has been elaborated and written in honor of Andrzej Bialas with whom I have collaborated during so many years and whom I consider both as a master and a close friend. Many domains of my research, including the dipole formalism (remembering exciting after-dinner sessions in our neighboring homes in the mid of 90's!), found their starting point and developments  in our lively discussions and our studies in common. for the dipole models see among others \cite{bialas}.

\end{document}